\documentclass[%
 aip,
 amsmath,amssymb,
 reprint,%
]{revtex4-1}

\usepackage{graphicx}
\usepackage{dcolumn}
\usepackage{bm}

\usepackage[utf8]{inputenc}
\usepackage[T1]{fontenc}
\usepackage{mathptmx}
\usepackage{etoolbox}
\renewcommand{\thesubsection}{(\arabic{subsection})}

\makeatletter
\def\@email#1#2{%
 \endgroup
 \patchcmd{\titleblock@produce}
  {\frontmatter@RRAPformat}
  {\frontmatter@RRAPformat{\produce@RRAP{*#1\href{mailto:#2}{#2}}}\frontmatter@RRAPformat}
  {}{}
}%
\makeatother
\begin{document}

\preprint{AIP/123-QED}

\title[A Noise-Robust Data Assimilation Method for Crystal Structure Prediction Using Powder Diffraction Intensity]{A Noise-Robust Data Assimilation Method for Crystal Structure Prediction Using Powder Diffraction Intensity}
\author{Seiji Yoshikawa}
\affiliation{Department of Physics, University of Tokyo, 7-3-1 Hongo, Bunkyo-ku, Tokyo 113-0033, Japan}

\author{Ryuhei Sato}%
\email{ryuhei.sato.c1@tohoku.ac.jp}
\affiliation{Department of Physics, University of Tokyo, 7-3-1 Hongo, Bunkyo-ku, Tokyo 113-0033, Japan}
\affiliation{Advanced Institute for Materials Research (WPI-AIMR), Tohoku University, 2-1-1 Katahira, Aoba-ku, Sendai 980-8577, Japan}

\author{Ryosuke Akashi}
\affiliation{Department of Physics, University of Tokyo, 7-3-1 Hongo, Bunkyo-ku, Tokyo 113-0033, Japan}

\author{Synge Todo}
\affiliation{Department of Physics, University of Tokyo, 7-3-1 Hongo, Bunkyo-ku, Tokyo 113-0033, Japan}
\affiliation{Institute for Solid State Physics, University of Tokyo, 5-1-5 Kashiwanoha, Kashiwa, Chiba 277-8581, Japan}
\affiliation{Mathematics and Informatics Center, University of Tokyo, 7-3-1 Hongo, Bunkyo-ku, Tokyo 113-8656, Japan}
\affiliation{Institute for Physics of Intelligence, University of Tokyo, 7-3-1 Hongo, Bunkyo-ku, Tokyo 113-0033, Japan}

\author{Shinji Tsuneyuki}
\affiliation{Department of Physics, University of Tokyo, 7-3-1 Hongo, Bunkyo-ku, Tokyo 113-0033, Japan}
\affiliation{Mathematics and Informatics Center, University of Tokyo, 7-3-1 Hongo, Bunkyo-ku, Tokyo 113-8656, Japan}
\affiliation{Institute for Physics of Intelligence, University of Tokyo, 7-3-1 Hongo, Bunkyo-ku, Tokyo 113-0033, Japan}
\affiliation{UTokyo Research Institute for Photon Science and Laser Technology, 7-3-1 Hongo, Bunkyo-ku, Tokyo 113-0033, Japan}

\date{\today}

\begin{abstract}
Crystal structure prediction for a given chemical composition has long been a challenge in condensed-matter science. We have recently shown that experimental powder X-ray diffraction (XRD) data are helpful in a crystal structure search using simulated annealing, even when they are insufficient for structure determination by themselves (N. Tsujimoto et al., Phys. Rev. Materials 2, 053801 (2018)). In the method, the XRD data are assimilated into the simulation by adding a penalty function to the physical potential energy, where we used a crystallinity-type penalty function defined by the difference between experimental and simulated diffraction angles. To improve the success rate and noise robustness, we introduce a correlation-coefficient-type penalty function adaptable to XRD data with significant experimental noise. We apply the new penalty function to SiO$_2$ coesite and $\epsilon$-Zn(OH)$_2$ to determine its effectiveness in the data assimilation method.
\end{abstract}

\maketitle

\section{\label{sec:level1}Introduction}
The prediction of crystal structures at the atomic level is one of the most fundamental challenges in condensed matter science\cite{c1}\cite{c2}. From the viewpoint of theoretical calculations, one problem is finding the global minimum of the potential-energy hypersurface. Thus, a number of methods have been developed to generate new candidate structures (ex. random sampling\cite{c3}, genetic algorithm\cite{c4,c5,c6,c7,c8,c9},  and particle swarm optimization (PSO)\cite{c10,c11,c12,c13}) and to overcome the potential barrier on the way to the target structure (ex. simulated annealing (SA)\cite{c14}, basin hopping\cite{c15}, minima hopping\cite{c16,c17},  and metadynamics\cite{c18}). Although these methods successfully yield the new structure represented by superconducting hydrides\cite{c8,c9,c19,c20,c21} and other materials\cite{c22,c23,c24}, they can only handle systems with up to a few hundred atoms because the number of candidate structures increases exponentially with system size. Therefore, it is necessary to accelerate these structure search methods when applied to the large systems.

One approach to accelerating the theoretical structure search is the utilization of the experimental data with conventional calculations\cite{c13,c25,c26,c27,c28,c29}. A typical approach is to calculate physical properties and directly compare them with experimental data to choose the best structure among the candidates\cite{c13,c26}. Another\cite{c25,c27,c28,c29} is to optimize the crystal structure in accordance with the cost function, which is the combination of the interatomic potential and the kind of "score" (penalty function) obtained from the similarity between the calculated and experimental physical properties during these simulations. In these data-assimilation techniques, we can predict structures from the experimental data, even if we do not know the crystal symmetry or do not have any candidate structures as initial configurations. Also, we can easily exclude the other stable structures, whose physical property does not agree with the experiment, since the structure is optimized in accordance with the cost function (i.e., interatomic potential + penalty function) taking into account the experimental data.

Putz et al.\cite{c28} confirmed that crystal structures such as TiO$_2$(rutile) and SiO$_2$(quartz) can be obtained from random structures by the Monte-Carlo method based on the linear combination of the empirical two-body potential and penalty function calculated from the similarity between experimental and calculated X-ray diffraction (XRD) patterns. Tsujimoto et al.\cite{c25}  showed that short-time (5000 step ($\approx$5 ps)) simulated annealing based on XRD-data-assimilated molecular dynamic (MD) simulation can even yield a low-symmetry structure such as SiO$_2$ coesite, although the success rate of finding the target structure is low (up to 30\% with an ideal experimental XRD pattern without noise). One of the problems in the previous study\cite{c25}  was the robustness against noise, since we need to manually determine whether each peak in the experimental data is the signal or noise by introducing a cutoff value as a pretreatment of experimental data.

Under such circumstances, we introduce a new correlation-coeﬃcient-type penalty function to the experimental (XRD) data-assimilated approach to improve the success rate and robustness to the experimental noise. We discuss whether this approach is applicable for hydrogen-containing material, in which hydrogen atoms scarcely contribute to the XRD pattern, to determine the structure.

\section{\label{sec:level2}Method}
\subsubsection*{A. Experimental data assimilation}
The central idea of the data assimilation method is based on Bayes' theorem. This method can be interpreted as a kind of maximum likelihood estimation. From Bayes' theorem, the conditional probability of the crystal structure given the experimental data is
\begin{equation}
\rho(\bm R{|}I_{\rm{ref}})=\frac{\rho(\bm R)\times\rho(I_{\rm{ref}}{|}\bm R)}{\rho(I_{\rm{ref}})}\rm{,} 
\end{equation}
where $\bm R$ represents the atomic positions and lattice parameters and $I_{\rm{ref}}$ is the reference experimental data. In this work, we adopt the diffraction pattern as $I_{\rm{ref}}$, though it can be arbitrary experimentally observable quantities. In the case of optimization by simulated annealing, the probability of finding the crystal structure follows the Boltzmann distribution:
\begin{equation}
\rho(\bm R)\propto \exp \{ -\beta E(\bm R) \} \rm{,} 
\end{equation}
where $E(\bm R)$ is the interatomic potential energy and $\beta$ is the inverse temperature. Meanwhile, by presuming that $\bm R$ giving calculated data ($I_{\rm{calc}}$) similar to $I_{\rm{ref}}$ should resemble the target structure, $\rho(I_{\rm{ref}}{|}\bm R)$ is defined in the following manner:
\begin{equation}
\rho(I_{\rm{ref}}{|}\bm R)\propto \exp \{ -\beta \alpha ND[I_{\rm{ref}},I_{\rm{calc}}(\bm R)] \}\rm{,}    
\end{equation}
where $N$ is the number of atoms and $\alpha$ is a weight parameter. $D$ is the penalty function obtained from the similarity between calculated and experimental data: $D$ takes zero when $I_{\rm{ref}}$ and $I_{\rm{calc}}$ match perfectly, whereas it takes positive value otherwise. As a result, from eqs.(1)-(3), $\rho(\bm R{|}I_{\rm{ref}})$ is proportional to the exponential of the cost function $F$ (eq.(5)) in the following manner:
\begin{equation}
\rho(\bm R{|}I_{\rm{ref}})\propto \exp \{ -\beta F(\bm R ; I_{\rm{ref}}) \}  \rm{,} 
\end{equation}
\begin{equation}
F(\bm R ; I_{\rm{ref}})=E(\bm R) + \alpha ND[I_{\rm{ref}},I_{\rm{calc}}(\bm R)]\rm{.} 
\end{equation}
Therefore, by minimizing $F$, we can maximize $\rho(\bm R{|}I_{\rm{ref}})$ to find the target structure referring to the given experimental data and efficiently obtain the target structure with this data-assimilated approach.
The concept of our method is summarized in the schematic diagram shown in Fig. 1. The target structure is the common minimum for both the interatomic potential energy and the penalty function. If the target structure is metastable or the experimental data is incomplete, the correct structure may not be the common global minimum of the potential energy and penalty function, yet it should be the best compromise as the global minimum of the cost function, $F$. In the case of XRD and Neutron diffraction patterns, the penalty function is mainly dependent on the symmetry and long-range order of the structure, while the interatomic potential mainly reflects the local interaction between atoms. Therefore, the positions of their local minima are different from each other and these local minima are expected to have significantly higher cost or become unstable by these two functions. As a result, the correct structure is more emphasized in the cost function than in the interatomic potential energy. Therefore, by optimizing the cost function, we can find the correct structure more efficiently than optimizing the structure with only the interatomic potential energy. 

\begin{figure}
\includegraphics[width=\linewidth]{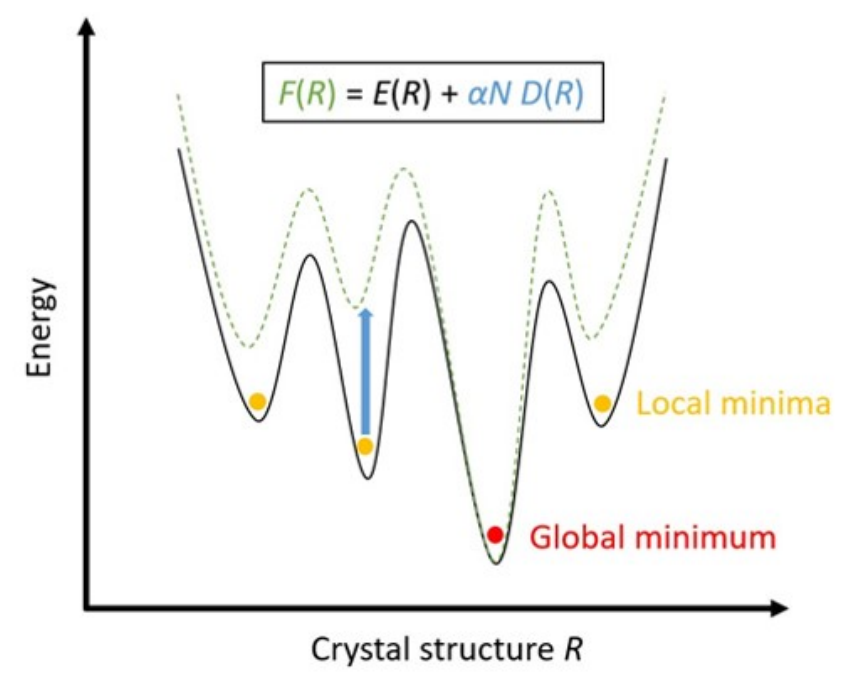}
\caption{\label{fig:epsart} Schematic image of the cost function. The horizontal axis represents the crystal structure $R$, which is actually a multidimensional space defined by cell parameters and atomic coordinates. The vertical axis represents the energy, which is optimized during the structure search. The black line represents the interatomic potential energy $E$, and the red and yellow circles are its global minimum and local minima, respectively. The blue arrow represents the penalty function ($\alpha ND(\bm R)$) added to the potential energy, and the green dotted line represents the cost function ($F=E(\bm R) + \alpha ND(\bm R)$).}
\end{figure}

\subsubsection*{B. Penalty function using the powder diffraction pattern}
As the penalty function in the previous study\cite{c25}, we used the crystallinity-type penalty function defined as
\begin{equation}
D_{\rm{Cryst}}(\bm R)=1-\frac{\int_{\theta=\theta_{obs}}I_{\rm{calc}}d\theta}{\int_{\theta_{\rm{min}}}^{\theta_{\rm{max}}}I_{\rm{calc}}d\theta}\rm{,}  
\end{equation}
where $I_{\rm{calc}}$ is the calculated diffraction intensity, $\theta$ is the diffraction angle, $[\theta_{\rm{min}}; \theta_{\rm{max}}]$ is the reference angle range, and $\theta_{obs}$ is the peak position observed in the reference experimental diffraction pattern $I_{\rm{ref}}$. This penalty function does not depend on the experimental intensity information and represents the degree of coincidence of peak positions between experimental and calculated diffraction patterns. By minimizing this penalty function, we can restrict the search space to satisfy the same extinction rule as in the experiment. Because of experimental errors such as the effect of preferred orientation, the peak intensity rate in experiments sometimes deviates from the ideal one. In such a case, this penalty function works well even with an unreliable experimental intensity rate. However, it does not work well when the target structure has low symmetry without the extinction law, or when it is difficult to determine peak positions owing to noise.

In this study, we propose to use the correlation coefficient between $I_{\rm{calc}}$ and $I_{\rm{ref}}$ as follows:
\begin{equation}
D_{\rm{CC}}(\bm R)=1-\frac{\int_{\theta_{\rm{min}}}^{\theta_{\rm{max}}} \tilde{I}_{\rm{calc}}(\bm R)\tilde{I}_{\rm{ref}} d\theta}{\sqrt{\int_{\theta_{\rm{min}}}^{\theta_{\rm{max}}}\tilde{I}_{\rm{calc}}^2(\bm R)d\theta}\sqrt{\int_{\theta_{\rm{min}}}^{\theta_{\rm{max}}}\tilde{I}_{\rm{ref}}^2d\theta}}\rm{,} 
\end{equation}
where $\tilde{I} = I - \bar{I}$ and $\bar{I}$ is the average intensity over the reference angle range $[\theta_{\rm{min}}; \theta_{\rm{max}}]$. This penalty function represents how well the shape of the calculated diffraction pattern agrees with that of the experiment. Therefore, it mainly fits the large peaks and their intensity rates in the experiment and includes not only the peak position but also the intensity information. The other feature of this penalty function is that it rarely requires pretreatment of experimental data. Even when the experimental noise is so large that we cannot distinguish some small diffraction peaks from noise, background subtraction is sufficient to capture the feature of the experimental XRD pattern using the correlation-coefficient-type penalty function. 

\subsubsection*{C. XRD-assimilated Molecular Dynamic simulation}
\begin{figure}
\includegraphics[width=\linewidth]{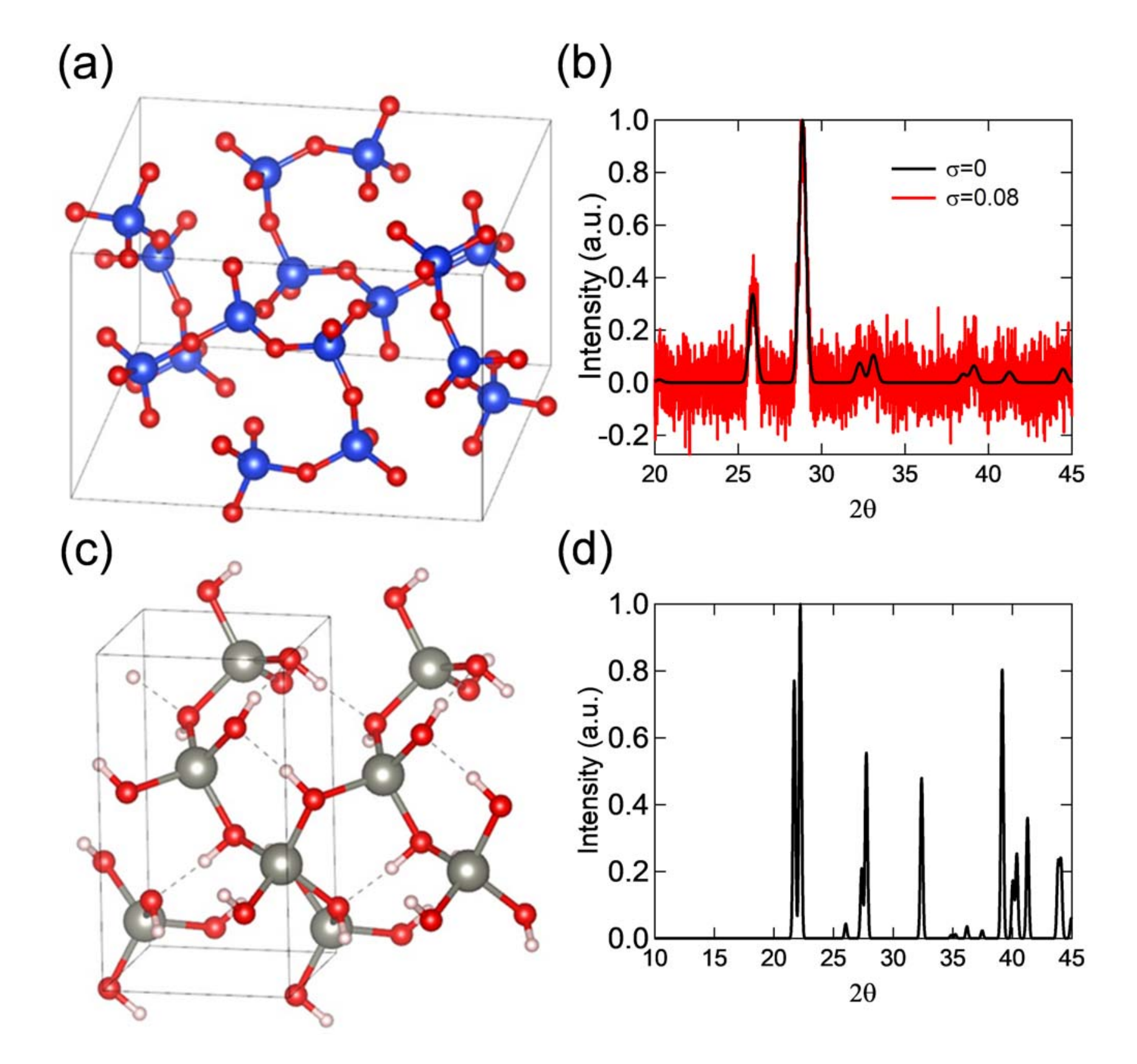}
\caption{\label{fig:epsart}Crystal structures of (a) coesite and (c) $\epsilon$-Zn(OH)$_2$. The blue and red spheres represent silicon and oxygen atoms in Fig. 1(a), while grey, red and pink spheres represent zinc, oxygen and hydrogen atoms in Fig. 1(c), respectively. The primitive cell for coesite ($a=7.14, b=12.37, c=7.12$ \AA, $\beta=119.57^{\circ}$) contains 48 atoms, while that for $\epsilon$-Zn(OH)$_2$ ($a=4.87, b=5.06, c=8.75$ \AA) contains 20 atoms. (b) Calculated (reference) XRD diffraction pattern smeared by Gaussian function with standard deviation of 0.2 degrees without (black) and with Gaussian white noise (red), whose standard deviation ($\sigma$) is 0.08. The wavelength $\lambda$ is set to 1.54 \AA ($\rm{Cu}$$K\alpha$-radiation). The reference range of the diffraction angle $[2\theta_{\rm{min}}; 2\theta_{\rm{max}}]$ is set to [20; 45] degrees. (d) Reference XRD patterns for $\epsilon$-Zn(OH)$_2$. $\lambda$ for XRD is set to 1.54 \AA. The diffraction peaks are smoothed with gaussian function with standard deviation of 0.1 degrees for XRD. The reference range of the diffraction angle $[2\theta_{\rm{min}}; 2\theta_{\rm{max}}]$ for XRD is set to [10; 45] degrees.}
\end{figure}

In this study, we perform data assimilated MD simulations to predict SiO$_2$ (coesite) and $\epsilon$-Zn(OH)$_2$ in the following manner: First, we set the appropriate simulation cell size to reproduce the experimental data in the target system. Although we used the known lattice parameters ($a=7.14, b=12.37, c=7.12$ \AA, $\beta=119.57^{\circ}$ for coesite: $a=4.87, b=5.06, c=8.75$ \AA\quad for $\epsilon$-Zn(OH)$_2$) in this report, in practice, we can use excellent programs for Miller indexing and subsequent lattice parameter determination, such as Conograph\cite{c30}. We employed a $2\times1\times1$ cell containing 96 (coesite) and 40 atoms ($\epsilon$-Zn(OH)$_2$) for the following data-assimilated MD simulations, since the primitive cell is considered to be too simple to test the applicability and effectiveness of the data-assimilated MD simulation. After choosing the lattice parameter, we generate random structures as initial atomic coordinates for each simulation. A structure with too small an interatomic distance is inappropriate as the initial arrangement, because the interatomic potential energy calculation sometimes diverges. Hence, to avoid the divergence of potential energy calculation during these simulations, atomic positions are chosen randomly, but the minimum interatomic distance is larger than 1.7 \AA\quad for coesite and 1.5 \AA\quad for $\epsilon$-Zn(OH)$_2$, when we generate the initial atomic coordinates. Then, we minimize the cost function by SA using MD simulation. Note that coesite is chosen for its low symmetry and $\epsilon$-Zn(OH)$_2$ as a test sample of a hydrogen containing structure. Also note that other SiO$_2$ structures such as alpha quartz and cristobalite can be determined by the same method, as shown in the previous study\cite{c25}.

For MD simulations, LAMMPS\cite{c31} packages were employed. During a 10,000-step (10ps) NVT simulation, the temperature is decreased from 10,000 (5,000 for $\epsilon$-Zn(OH)$_2$) to 0 K linearly by the velocity scaling method. For calculating the total energy and forces, we adopted the Tsuneyuki potential\cite{c32} for SiO$_2$, whereas we executed the DFT calculation (VASP\cite{c33,c34}) for $\epsilon$-Zn(OH)$_2$. The total energy is examined to determine whether the correct structure has been obtained. In DFT calculation, the Perdew--Burke--Ernzerhof generalized gradient approximation (PBE-GGA)\cite{c35} and projector augmented wave (PAW) method\cite{c36,c37} were employed for the exchange-correlation function and pseudopotential, respectively. The plane-wave energy cutoff was set to 400 eV and only $\Gamma$ point was sampled for the Brillouin zone integration. The reference diffraction data in Figs. 2(b) and (d) are calculated from the experimentally known structure in Figs. 2(a) and (c). Here, as in the previous work\cite{c25}, we intentionally limit the range of diffraction angles to 20-45 degrees for coesite and 10-45 degrees for $\epsilon$-Zn(OH)$_2$. One reason is that the peak intensity decreases with increasing diffraction angle due to the XRD atomic scattering factor and the Debye-Waller factor, making it difficult to separate the peak intensity from the noise. Another reason is that we cannot measure large-angle diffraction in ultrahigh-pressure experiments where the data-assimilation method is highly required. Gaussian white noise is also introduced to the reference diffraction data (Fig. 2(b)) to examine the noise robustness of the data assimilation method.

\section{Results and discussion}
\subsubsection*{A. Correlation-coefficient-type penalty function}
The success rate of finding the target structure is significantly improved by using the new correlation-coefficient-type penalty function. Figure 3(a) shows the success rate of finding the coesite structure using XRD-assimilated MD simulations as a function of the weight parameter $\alpha$. Note that the obtained structure is regarded as that for coesite when its interatomic potential is about --1716.78 eV. We performed 50 simulations for each value of $\alpha$ and with two types of penalty function. The weight parameter $\alpha$ represents how strongly the search space is restricted by the penalty function so that the calculated XRD pattern matchs that of the experiment. If $\alpha$ is set to zero, it means that MD simulations have been performed using only the interatomic potential energy without the experimental data. As shown in the figure, with appropriate $\alpha$, the success rate of finding the coesite structure almost exceeds 80\% using the correlation coefficient ($D_{\rm{CC}}$). On the other hand, the success rate obtained using the crystallinity ($D_{\rm{Cryst}}$) is at most 50\%. It is also found that the success rate for the simulations without the penalty function ($\alpha$=0) is about 0\%, showing that the efficiency of crystal structure search is enhanced by the penalty function. 
\begin{figure}
\includegraphics[width=\linewidth]{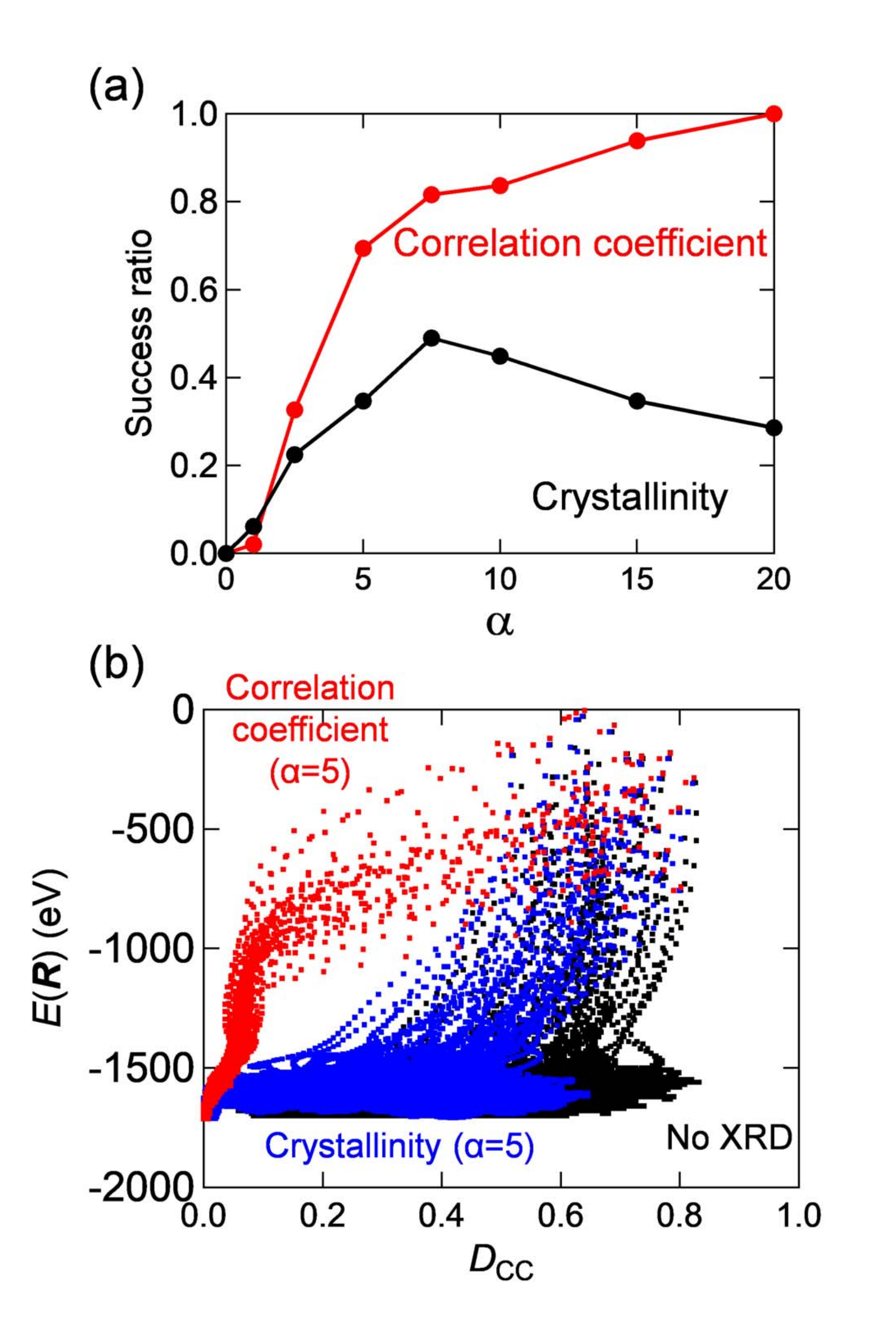}
\caption{\label{fig:epsart} (a) Success rates of finding coesite structure using correlation-coefficient-type and crystallinity-type penalty functions. The horizontal axis $\alpha$ is the weight parameter in eq. (5). (b) Interatomic potential energies and penalty function, $D_{\rm{CC}}$, values for the trajectories of 50 data-assimilated MD simulations targeting SiO$_2$ coesite. Data obtained by the data-assimilated MD simulations with crystallinity-type (blue) and  correlation-funciton-type (red) penalty functions and without (black) penalty function.}
\end{figure}

The above conclusion is further confirmed by the interatomic potential energy of the optimization trajectories as a function of $D_{\rm{CC}}$ (Fig.3(b)). This figure shows the values of interatomic potential energy and $D_{\rm{CC}}$ from 0 to 10,000 steps for all 50 data-assimilated MD simulations. According to the definition of the correlation-coefficient-type penalty function, the calculated XRD pattern is identical to the reference pattern when $D_{\rm{CC}}$ = 0, whereas  the calculated XRD pattern is totally different from the reference when $D_{\rm{CC}}$=1. We evaluated $D_{\rm{CC}}$ for the three optimization cases, ``No XRD", ``Crystallinity" and ``Correlation coefficient", in order to compare the distribution of the structural similarity. Here we note that, for the former two cases, $D_{\rm{CC}}$ was not referred to for the optimization. Without the XRD penalty function, most of the structures obtained during MD simulations have small energies as low as --1700 eV, but $D_{\rm{CC}}$ is widely distributed from 0.06 to 0.8, meaning that this material has many metastable structures and tends to become amorphous. On the other hand, by introducing the correlation-coefficient-type penalty function, most of these configurations become unstable and are rarely obtained during XRD-assimilated MD simulation. Therefore, the structures with an energy of about –-1700 eV during the XRD-assimilated MD simulation are distributed only around $D_{\rm{CC}}$=0.0. Compared with the correlation coefficient, the crystallinity weakly restricts the search space, as shown in the distribution around --1700 eV, which has wide ranging $D_{\rm{CC}}$ from 0.0 to 0.6. Thus, we conclude that the correlation-coefficient-type penalty significantly accelerates the structure search compared with the previous crystallinity-type penalty.

\subsubsection*{B. Diffraction pattern with noise}
The use of correlation coefficient improves not only the success rate but also the noise robustness. To verify the noise robustness of the data-assimilated structure prediction method, we add Gaussian white noise to the reference XRD pattern, as shown by the red line in Fig. 2(b). In this figure, the standard deviation of the noise, $\sigma$, is set to 0.08 times the height of the maximum diffraction peak. The intensity with noise at each $\theta$ is calculated using the reference intensity ($\sigma$=0) in the following manner:
\begin{equation}
I(\theta)=I(\theta)_{\sigma=0} + \xi  I_{\rm{MAX},\sigma=0}\rm{.} 
\end{equation}
Here, $I_{\rm{MAX},\sigma=0}$ and $\xi$ represent the intensity of the largest diffraction peak and a random number, respectively. The random number is chosen so that its probability distribution equals to Gaussian distribution function with $\sigma$:
\begin{equation}
P(\xi)= \frac{1}{\sqrt{2\pi\sigma^2}} \exp(-\frac{\xi^2}{2\sigma^2})\rm{.} 
\end{equation}
As shown in the figure, because of the large noise, it is almost impossible to distinguish the signal from the noise except for the largest and second largest peaks at 25.9 and 28.9$^\circ$, respectively. Figure 4 shows the success rate of finding the coesite structure using XRD-assimilated MD simulations with the reference XRD pattern including Gaussian white noise. We perform 50 simulations for each value of $\alpha$ and amplitude of noise $\sigma$. According to eq.(8) and (9), the amplitude of noise is roughly proportional to the standard deviation ($\sigma$) for the Gaussian white noise. Compared with the case for the ideal reference XRD pattern without noise ($\sigma$=0), the success rate decreases as the amplitude of noise increases. However, even with the largest $\sigma$ in these test simulations ($\sigma$=0.12), the success rate is about 33\% with appropriate $\alpha$, showing the noise robustness of the correlation-coefficient-type penalty function. 

\begin{figure}
\includegraphics[width=\linewidth]{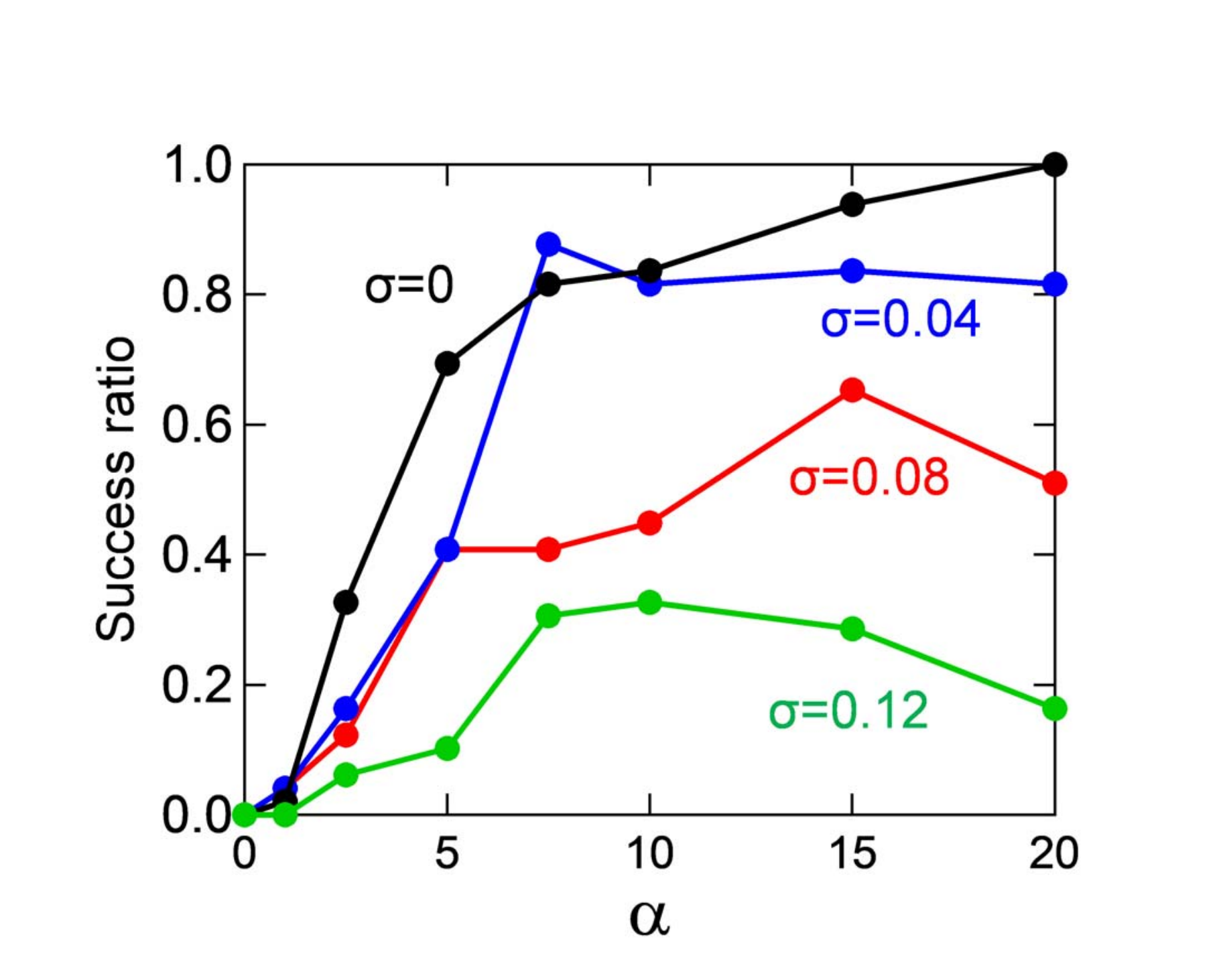}
\caption{\label{fig:epsart} Success rate of finding coesite structure using XRD-assimilated MD simulations with noisy reference XRD data as a function of the weight parameter $\alpha$. $\sigma$ represents the standard deviation for the introduced Gaussian white noise.}
\end{figure}

\subsubsection*{C. System containing hydrogen}
The applicability of data-assimilated MD simulations to hydrogen containing materials is discussed in this section. Figure 5 shows the interatomic potential energy distribution obtained by XRD-assimilated MD simulations as a function of $D_{\rm{CC}}$. Note that, for $\epsilon$-Zn(OH)$_2$, various structures hosting different hydrogen bonding or H$_2$O molecules also have comparably small $D_{\rm{CC}}$. To highlight the  subtle difference in $D_{\rm{CC}}$, we plotted $\log D_{\rm{CC}}$ in the horizontal axis. As shown in the figure, without the XRD penalty function, there is a local minimum at around log($D_{\rm{CC}}$)= –0.9-0.0, suggesting that conventional SA using DFT-MD simulation rarely leads to the $\epsilon$-Zn(OH)$_2$ structure. On the other hand, XRD-assimilated DFT-MD simulation can yield the $\epsilon$-Zn(OH)$_2$ structure, as shown by the structures obtained at around $\log D_{\rm{CC}}$ = –5 (purple dots in the figure). However, there are still local minima at around $\log D_{\rm{CC}}$=–2.35 and –4, which reduce the success rate. These local minimums are probably related to H$_2$O molecule formation, since almost all of these structures at $\log D_{\rm{CC}}$=–2.35 and –4 (purple dots in Fig. 5) include H$_2$O molecules. Note that 97\% of the structures obtained after DFT-MD simulation without the XRD penalty function include H$_2$O molecules, suggesting that the XRD penalty function prevents H$_2$O molecule formation to some extent, although 70\% of the structures obtained by XRD-assimilated DFT-MD simulations still include H$_2$O molecules. Therefore, introducing some constraint to reduce H$_2$O molecule formation during data-assimilated MD simulations could further increase the success rate. Nevertheless, the success rate of finding the $\epsilon$-Zn(OH)$_2$ structure is 30\% (the rate of finding structures whose interatomic potential energy is about -190.3 eV\cite{c38}) with the XRD penalty function with weight parameter $\alpha$=15, showing the effectiveness of XRD-assimilated simulation in case of the hydrogen-containing system. 

\begin{figure}
\includegraphics[width=\linewidth]{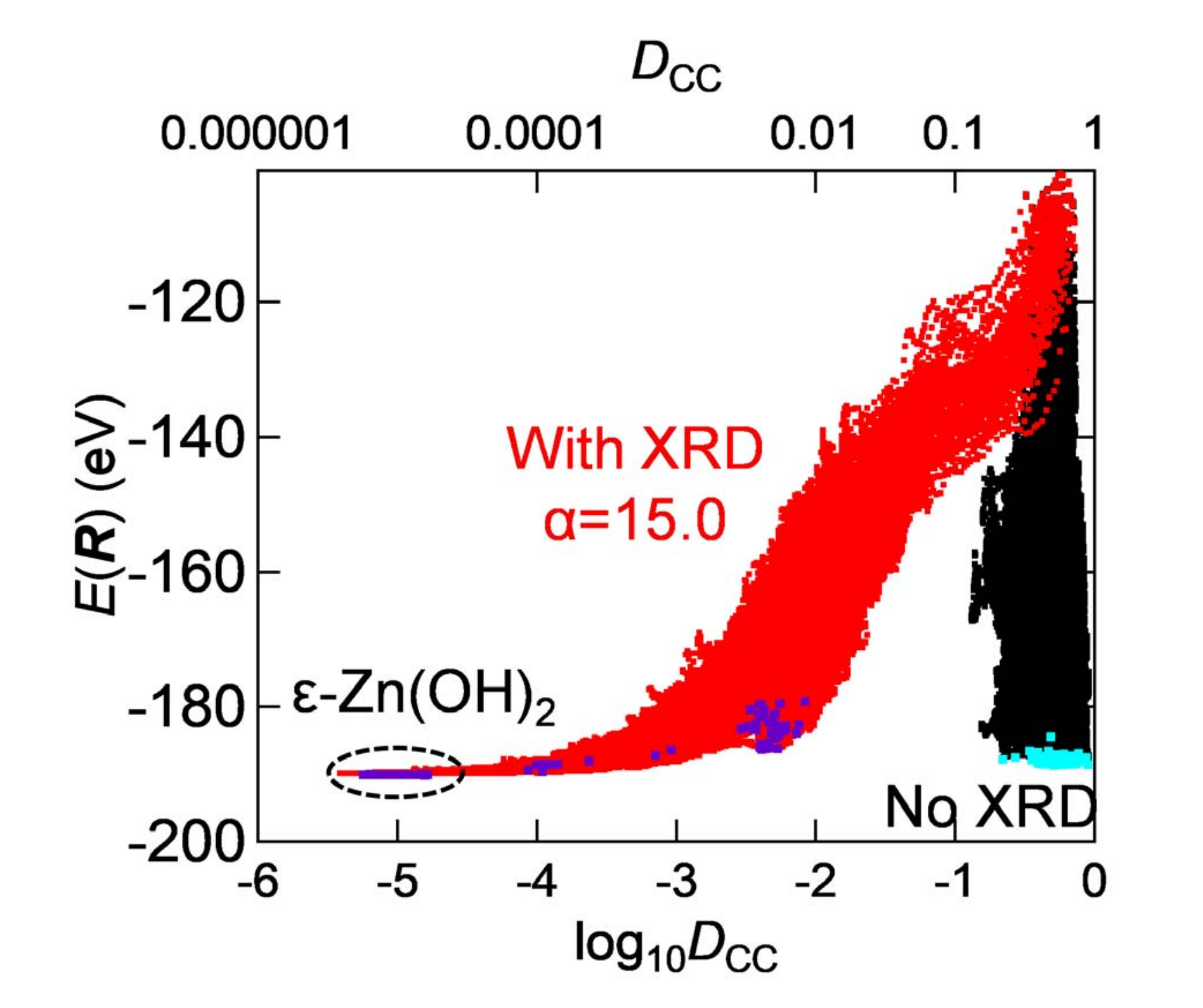}
\caption{\label{fig:epsart} Interatomic potential energies and penalty function $D_{\rm{CC}}$ values for the trajectories of 100 data-assimilated MD simulations targeting $\epsilon$-Zn(OH)$_2$ (red) and those of ordinary MD simulations. The purple and cyan dots show the interatomic potential energies of the obtained structures after the data-assimilated and ordinary MD simulations. }
\end{figure}

\section{Conclusion}
In this study, we introduce a new correlation-coefficient-type penalty function for XRD data-assimilated simulations to improve the success rate of finding the target structure and noise robustness. Using the correlation-coefficient-type penalty function, the success rate of finding the coesite structure with noise-free reference data almost exceeds 80\%, which is almost twice as much as that obtained with the crystallinity-type penalty function. In addition, the use of this penalty function enable us to predict the coesite structure even with noisy reference data in which the minor peaks are almost indistinguishable from the noise. It is also found that this XRD data-assimilated MD simulation can be applied to the hydrogen-containing materials even though the hydrogen atoms negligibly contribute to the XRD data. The success rate of finding $\epsilon$-Zn(OH)$_2$ structure is about 30\% using the reference XRD pattern and DFT potential, suggesting that this method is useful for the structure determination of high-pressure metal hydrides, where the angle range of XRD patterns is too small to determine the structure through an experimental approach.

\begin{acknowledgments}
This work was supported by JSPS KAKENHI Grant-in-Aid for Scientific Research on Innovative Areas "Hydrogenomics", No. JP18H05519, and Elements Strategy Initiative to Form Core Research Center in Japan. S.Y. was supported by the Japan Society for the Promotion of Science through the Program for Leading Graduate Schools (MERIT).
\end{acknowledgments}

\section*{AUTHOR DECLARATIONS}
\section*{Conflict of Interest}
The authors have no conflicts to disclose.
\section*{Data Availability}
The data that support the findings of this study are available from the corresponding author upon reasonable request.

\section*{Reference}
\nocite{*}
\bibliography{aipsamp}

\begin{thebibliography}{38}
\bibitem [{c38()}]{c38}%
  \BibitemOpen
  \href@noop {} {\ }\bibinfo {note} {The potential energy for the obtained
  $\epsilon$-Zn(OH)$_2$ structures ($\approx$ –190.3 eV) is slightly larger
  than that of reference $\epsilon$-Zn(OH)$_2$ (–190.4 eV). This suggests
  that the hydrogen bond orientation in the obtained structure does not
  perfectly agrees with that of reference. However, this energy difference is
  considered to be within the range of the error, because this energy
  difference (2.5 meV/atom = 0.1 eV for 40 atoms) is much smaller than thermal
  energy at 300K (=25 meV/atom = 1 eV for 40 atoms). In addition, in terms of
  XRD pattern, we can hardly distinguish the obtained structure from reference
  one, since $D_{CC}$ is about $10^{-5}$. Therefore, here, we regard the
  obtained structure with total energy at around –190.3 eV as
  $\epsilon$-Zn(OH)$_2$.}\BibitemShut {Stop}%
\end{thebibliography}

\end{document}